\documentclass[twocolumn,showpacs,amssymb,amsmath,nobibnotes,prl]{revtex4}

\usepackage{graphicx}
\usepackage{xcolor}
\usepackage{bm}%

\begin{document}
\title{Non-Dissipative Saturation of the Magnetorotational Instability in Thin Disks}
\author{Edward Liverts} \email{eliverts@bgu.ac.il}
\author{Yuri Shtemler} \email{yshtemler@gmail.com} \author{Michael Mond} \email{mond@bgu.ac.il}\affiliation{Department of Mechanical
Engineering, Ben-Gurion University of the Negev, Beer-Sheva 84105,
Israel}
\author{Orkan M Umurhan} \email{oumurhan@ucmerced.edu}
\affiliation{School of Natural Sciences, UC Merced, Merced, CA 95343, USA, and\\
City College of San Francisco, San Francisco, CA 94112, USA}
\author{Dmitry V Bisikalo}\email{bisikalo@inasan.rssi.ru}
\affiliation{Institute of Astronomy of the Russian Academy of Science, 48 Pyatnitskaya, Moscow, Russia}
\date{\today}

\begin{abstract}
A new non-dissipative mechanism is proposed for the saturation of the axisymmetric magnetorotational (MRI) instability in thin Keplerian disks that are subject to an axial magnetic field. That mechanism relies on the energy transfer from the MRI to stable magnetosonic (MS) waves. Such mode interaction is enabled due to the vertical stratification of the disk that results in the discretization of its MRI spectrum, as well as by applying the appropriate boundary conditions. A second order Duffing-like amplitude equation for the initially unstable MRI modes is derived. The solutions of that equation exhibit bursty nonlinear oscillations with a constant amplitude that signifies the saturation level of the MRI. Those results are verified by a direct numerical solution of the full nonlinear reduced set of thin disk magnetohydrodynamics equations.
\end{abstract}

\pacs{47.65.Cb, 43.35.Fj, 62.60.+v}
\maketitle

\textit{Introduction - }
The destabilizing effect of an axial magnetic field on Couette flow has been discovered half a century ago {\cite{Velikhov}, \cite{Chandra}. However, the importance of that phenomenon to astrophysics has
been recognized only three decades later by Balbus and Hawley \cite{Balbus1}, \cite{Balbus3}. That mechanism, termed the magnetorotational instability
(MRI), is indeed considered by many researchers as the main candidate to hold the key to solving the problem of
angular momentum transfer in accretion disks, and has been thoroughly investigated through linear analysis
as well as nonlinear magnetohydrodynamic (MHD) simulations under a wide range of conditions and
applications. Because of its importance in accretion disks physics the understanding of the MRI's saturation mechanisms and level is of utmost significance. Thus, Knobloch \& Julien \cite{KnoblochJulien} have described analytically the nonlinear saturation of the MRI in a straight infinite vertically uniform channel with solid boundaries. Such configuration is characteristic of laboratory experiments rather than astrophysical conditions. Knobloch \& Julien considered a developed stage of the MRI, far from its threshold and showed that as the presence of the solid boundaries supports radial pressure gradients, the latter act together with the viscous as well as Ohmic dissipation in order to modify the rotation shear that feeds the instability and thus saturating it. In a complementary work, Umurhan et al. \cite{Umurhan et al} performed a weakly nonlinear analysis of the MRI close to marginality in configurations similar to \cite{KnoblochJulien} and showed that the MRI saturates due to dissipative effects to levels that scale with the square root of the magnetic Prandtl number. In the current letter a novel mechanism for the saturation of the MRI is proposed, that differs from the above two works in the following important two aspects: 1. A true thin disk is considered that is characterized by vertically localized stratified mass density,  subject to radiation boundary conditions, and 2. The proposed mechanism is non-dissipative. Indeed, it is shown that the nonlinear forcing of magnetosonic (MS) waves by the MRI results in the saturation of the latter, and leads to bursty nonlinear oscillation of its amplitude.\\
\textit{The Physical Model - } The thin disk asymptotic expansion procedure is applied to the MHD equations in order to study the weakly nonlinear evolution of the MRI.
The underlying physical property of the system is the supersonic nature of the Keplerian rotation whose Mach number is proportional to $1/\epsilon $,
where $\epsilon$ is the ratio of the disk's semi-thickness to its characteristic radius. Consequently, both steady-state as well as the perturbed variables are scaled with well defined powers of the small parameter $\epsilon$. Such procedure has been employed in numerous studies of thin disk dynamics, \cite{1}-\cite{5}, and has in particular been proven efficient in the realistic analysis of the discrete MRI spectrum in true thin disk geometry, \cite{6}. Thus, the steady-state configuration is characterized by a Keplerian rotation $\Omega(r)=r^{-3/2}$ where $\Omega(r)$ and $r$ are normalized by the value of the rotation frequency at some radius $R_0$, and $R_0$, respectively, as well as by an axial magnetic field $B_z(r)$ that is an arbitrary function of $r$. The axial steady-state structure of the disk is determined from the force balance between the thermal pressure and the gravitational pull of the central object. Thus, assuming axially isothermal configuration results in the following normalized number density profile: $n(r,\zeta)=N(r)\Sigma(\eta)$, where $\Sigma(\eta)=e^{-\eta ^2/2}$, $N(r)$ is an arbitrary function of $r$, $\eta =\zeta/H(r)$, $\zeta=z/\epsilon$ is the stretched axial coordinate, and $H(r)$ is the semi thickness of the disk. The latter [or alternatively the temperature profile $T(r)$] is an arbitrary function of $r$.

As the axial variations of the perturbations are assumed to occur on a much smaller length scales than the corresponding radial changes, to lowest order the MHD equations are given by the following set of \emph{reduced nonlinear equations} that depend parametrically on the radial coordinate (see \cite{6} for detailed derivation):
\begin{equation}
 \frac{\partial v_r}{\partial t}
-2v_\theta
-\frac{1}{\beta(r)}\frac{1}{\Sigma (\eta)+\sigma}\frac{\partial b_r}{\partial \eta }  =
-v_z\frac{\partial v_r}{\partial \eta},
\label{29}
\end{equation}
\begin{equation}
\frac{\partial v_\theta }{\partial t}
+\frac{1}{2}v_r
-\frac{1}{\beta(r)}\frac{1}{\Sigma (\eta)+\sigma}\frac{\partial b_\theta}{\partial \eta }
=-v_z\frac{\partial v_\theta}{\partial \eta},
\label{30}
\end{equation}
\begin{equation}
\frac{\partial b_r}{\partial t}
-\frac{\partial v_r}{\partial \eta}=
-\frac{\partial (v_z b_r)}{\partial \eta },
\label{31}
\end{equation}
\begin{equation}
\frac{\partial b_\theta}{\partial t}
-\frac{\partial v_\theta}{\partial \eta}
+\frac{3}{2} b_r=
-\frac{\partial (v_z b_\theta)}{\partial \eta }
,
\label{32}
\end{equation}
\begin{eqnarray}
\nonumber
\frac{\partial  v_z}{\partial t}
+\frac{\Sigma(\eta)}{\Sigma(\eta)+\sigma}
\frac{\partial}{\partial \eta }\bigl[\frac{\sigma}{\Sigma (\eta)}\bigr]
=\\
-\frac{1}{2}\frac{\partial v_z^2}{\partial \eta}
-\frac{1}{2\beta(r)[\Sigma (\eta)+\sigma]}\frac{\partial (b_r^2+b_\theta^2)}{\partial \eta},
\label{33}
\end{eqnarray}
\begin{equation}
\frac{\partial \sigma}{\partial t}
+\frac{\partial [\Sigma (\eta)+\sigma]v_z}{\partial \eta }=0,
\label{34}
\end{equation}
where time is scaled with the local inverse rotation frequency,  $\sigma$, $(v_r,v_{\theta},v_z)$ and $(b_r,b_{\theta},b_z)$ are the perturbed number density (scaled by $N(r)$) and the components of the perturbed velocity (scaled by the sound velocity, $C_s(r)$), and magnetic field (scaled by the steady-state axial magnetic field, $B_z(r)$,  respectively. In addition, the local plasma beta is given by $\beta(r)=\beta_0 N(r)C_s^2(r)/B_z^2(r)$ where $\beta_0$ is the beta value at $R_0$, and $C_s(r)$ is proportional to the square root of the temperature.  \\
\textit{Small Perturbations - }Linearizing the system of equations (\ref{29})-(\ref{34}) the latter decouples into two sub-systems: the first one is obtained from eqs. (\ref{29})-(\ref{32}) and describes the evolution of two Alfv\'{e}n-Coriolis waves one of which is responsible for the MRI. The second sub-system is obtained from eqs. (\ref{33})-(\ref{34}) and describes the MS modes. Liverts \& Mond \cite{LivertsMond} obtained a WKB solution for the MRI eigenvalues and eigenfunctions for Gaussian stratified disks. However,
modifying the steady state number density axial profile to ${\bar\Sigma} (\eta) =sech^2\eta$ enables the analytical solution of both sub-systems. Assuming that the perturbations evolve as $e^{i\lambda t}$, the eigenfunctions of the Alfv\'{e}n-Coriolis sub-system can be expressed in terms of the Legendre polynomials which leads to the following dispersion relation \cite{6}:
\begin{equation}
(3\beta_{cr}^k-\lambda ^2 \beta)[3\beta_{cr}^k-(3+\lambda ^2 )\beta]-4\lambda ^2 \beta ^2=0,
\end{equation}
 where $\beta _{cr}^k = k(k+1)/3, k=1,2,\ldots$. As can be inferred from eq. (7), due to the axial stratification of the steady-state configuration the unstable modes are quantized with the axial number $k$ which is  equivalent to the axial wave number for the axially uniform case. Of particular interest is the fact that the number of unstable MRI modes increases with $\beta$, the threshold for exciting $k$ unstable modes being $\beta _{cr}^k$. Figure 1 depicts the emergence of more and more unstable modes as $\beta$ is increased. It is further noticed that each point on the $\beta$-axis  with $\beta =\beta _{cr}^k, k=1,2,\ldots$, serves as a bifurcation point for two modes, namely an unstable MRI one with $+\gamma$ and a stable mode with $-\gamma$. Consequently, at the bifurcation points the eigenvalue $\lambda$ is zero with multiplicity 2. This fact will turn out to be of great significance in the weakly nonlinear analysis to be unfolded in the next sections.\\
Turning to the MS sub-system, its spectrum is stable and continuous. The eigenfunctions may be expressed as linear combinations of the following pair of linearly independent functions, \cite{7}: $f_\pm=[(1-\xi)/(1+\xi)]^{\pm\mu/2}(\mu\pm \xi)$, where $\xi = tanh\eta$, and $\mu = \sqrt{1-\lambda ^2}$. It can be easily proven that solutions that vanish at $\xi =\pm 1$ exist only for $\lambda ^2 >0$, which indeed renders the MS modes stable.

\begin{figure}[h]
\centering
\includegraphics*[width=60mm,height=45mm]{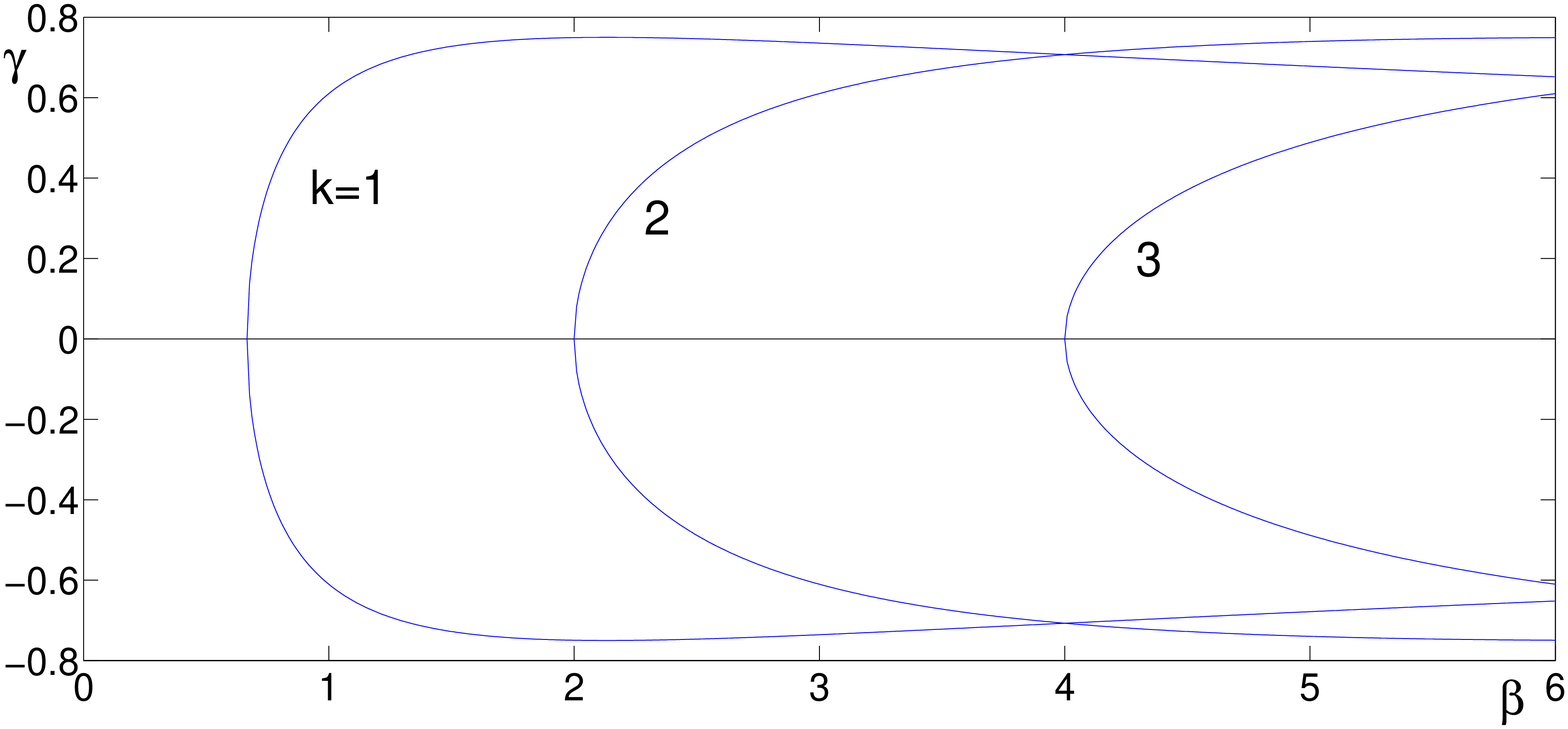}
\caption{The bifurcation diagram for MRI modes for k=1,2,3 as a function of $\beta$. Each MRI branch that is characterized by a mode number $k$ is accompanied by a stable branch that is symmetric about the vertical axis.}
\end{figure}
\textit{Weakly Nonlinear Analysis - } The focus is put now on the portion of the $\beta$-axis that is only slightly above $\beta_{cr}^1=2/3$. As can be seen from Fig. 1, which depicts the solution of eq. (7) for the first MRI modes, in that range of values there is only one such mode, namely, for $k=1$. Defining the control parameter $ \delta\ll 1$ in the following way:  $\beta=\beta_c^1+\delta$, the growth rate of the single unstable MRI mode may be calculated to lowest order in $\delta$ from eq. (7) to be $\gamma ^2 =27\delta/14$. The expression for the eigenfunction of, say, the radial component of the perturbed magnetic field of that initially small MRI perturbation looks, therefore as follows, \cite{7}:
\begin{equation}
b_r(r,\eta,t)=b_0(r)[P_0(\xi )-\xi P_1(\xi )]a(t),
\end{equation}
where $a(t)=a_0e^{\gamma (\delta )t}$, $P_k$ is the Legendre polynomial of order $k, k=0,1$, $\xi = tanh\eta$, $b_0$ is an arbitrary function of $r$, and $a_0$ is the amplitude of the initial perturbation.

The amplitude $a$ of the growing MRI remains exponential as long as the perturbation is small enough. However, as the perturbation grows exponentially, nonlinear effects kick in. Thus, it is clear from eqs. (5) and (6) that the pressure exerted by the perturbed magnetic field excites the MS modes which in turn alters the stability properties of the original MRI mode by affecting its axial convection as well as by modifying the Alfv\'{e}n velocity. As a result of that interaction of the growing MRI with the excited MS mode the amplitude of the former is no longer exponential but becomes a different function of time. The aim of the weakly nonlinear analysis is therefore to derive an appropriate ordinary differential equation that describes the evolution in time of the amplitude $a(t)$ of the single MRI mode. In order to do that it is recalled that the transition to instability (when $\beta$ reaches $\beta_{cr}^1$ from below) occurs when the linearized system of equations has a double zero eigenvalue. As a result the sought after equation is expected to be of second order as opposed to first order equations that characterize systems that bifurcate to instability through a simple zero eigenvalue \cite{8}. Thus, the appropriate amplitude equation is of the following form:
\begin{equation}
\frac{d^2 a}{dt^2}=\gamma ^2 a-\alpha a^3.
\end{equation}
 Equation (9) has been derived directly from eqs. (1)-(6) by suitable asymptotic analysis . However, for sake of brevity of the current presentation, $\alpha$  is determined below from the parameters that characterize the steady-state. Equation (9) has also been used by Arter \cite{9} in order to describe sawtooth oscillation in Tokamaks. The role of double eigenvalues has also been recognized by Stefani and Gerbeth \cite{Stefani} in order to study polarity reversals in mean-field dynamo models.

In order to calculate $\alpha$ it is first noticed that eq. (9) shares two important features with the full set of the reduced nonlinear equations (1)-(6): 1. For $\delta <0$ the origin of the phase space (i.e., the $a-da/dt$ plane) is the only fixed point of the dynamical system described by eq. (9). This is a reflection of the fact that for $\delta <0$ the only steady-state solution of the set (1)-(6) is the original basic Keplerian rotation. 2. For $\delta >0$ the single fixed point at the origin turns into a saddle point while two additional fixed points emerge which are centers and are located in symmetrically opposite locations with respect to the origin. The latter occur due to nonlinear effects and are given by $da_c/dt=0, a_c^2=\gamma ^2/\alpha$. This kind of behaviour, once again, reflects the fact that once the MRI grows exponentially, the nonlinear terms give rise to a new stable steady-state which may be calculated through eqs. (1)-(6). Having that in mind a clear strategy emerges for calculating the value of $\alpha$ , namely,  by the amplitude of the non linear steady state solution of eqs. (1)-(6).

%
%

\textit{The Nonlinear Steady-State - } The first steady-state solution of the set of equations (1)-(6) is obviously given by setting all the physical variables to zero. That solution describes the original basic Keplerian flow without any perturbations. However, as the beta value protrudes into the unstable region in parameter space, an additional \textit{perturbed} steady-state solution emerges. Finding that nonlinear steady state starts by realizing that $b_{\theta}^{ns}(\xi)=v_r^{ns}(\xi)=v_z^{ns}(\xi)=0$, where the superscript $ns$ denotes the nonlinear steady state solution. Consequently, the following single ordinary differential equation is derived for the radial component of the steady state magnetic field:
\begin{equation}
\frac{d^2b_r^{ns}}{d\xi ^2}+3\beta\frac{b_r^{ns}}{1-\xi ^2}+\frac{d}{d\xi}\Big [\frac{(b_r^{ns})^2}{\beta (1-\xi ^2)-(b_r^{ns})^2}\frac{db_r^{ns}}{d\xi}\Big ]=0,
\end{equation}
where, as written above, $\beta = \beta _{cr}^{1} +\delta$ with $\beta _{cr}^{1}=2/3$. An asymptotic solution of eq. (10) in the limit of small positive $\delta$ is now obtained by expanding  $b_r^{ns}$ in the following power series in $\sqrt{\delta}$ (or alternatively in powers series in $\gamma$):
\begin{equation}
b_r^{ns}(\xi)=\sqrt{\delta} \mu_1 b_r^{(1)}(\xi)+(\sqrt{\delta})^3 \mu_3b_r^{(3)}(\xi)+\ldots .
\end{equation}
To lowest order in $\sqrt{\delta}$ the solution is given by the right hand side of eq. (8), exactly as the first linear eigenfunction. The coefficient $\mu_1$ is determined now by the solvability condition of the resulting inhomogeneous equation for $b_r^{(3)}(\xi)$. Applying for that purpose Fredholm's alternative theorem yields: $\mu_1=\sqrt{5/2}$. Recalling now that the center fixed point of the dynamical system described by eq. (9) is given by $a_c^2=\gamma ^2/\alpha$, and that $a_c$ is identified with $\sqrt{\delta}\mu_1$, the value of $\alpha$ is readily computed to be $\alpha=2\gamma ^2/ 5 \delta$.

\begin{figure}[h]
\centering
\includegraphics*[width=60mm,height=45mm]{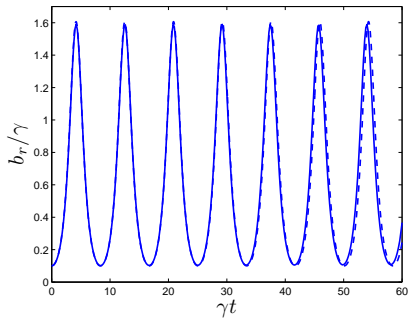}

\caption{Comparison between $b_r(\eta =0,t)$ as obtained from eqs. (1)-(6) (full line), and the solution of eq. (9) (dashed line), both for $\delta=0.0052$.}
\end{figure}
\begin{figure}[h]
\centering
\includegraphics*[width=60mm,height=45mm]{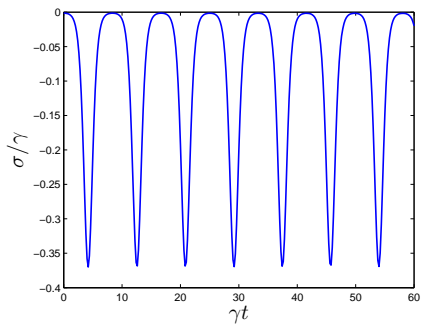}
\caption{The amplitude of the perturbed number density in the equatorial plane, i.e., $\sigma(\eta =0,t)$, as obtained from eqs. (1)-(6) for $\delta =0.0052$.}
\end{figure}

\textit{Results - } Equation (9) is known in the literature as the undamped Duffing equation and its solutions
may be expressed analytically in terms of the Jacobi elliptic functions. Of particular importance is the fact that that kind of Duffing equation is derivable from an Hamiltonian and hence its solutions are bounded. This reflects indeed the saturation of the MRI. For initial conditions near the saddle point at the origin of the phase plane the period of the bursty oscillations is asymptotically given by $\Theta \rightarrow ln(8/h)/2\gamma$ as $h \rightarrow 0$, where $h=\alpha  (\dot a ^2-\gamma ^2 a^2+ a^4/2)_{t=t_0}/4\gamma ^2$. Consequently, a small change in the initial conditions may dramatically change the period of the nonlinear oscillations without affecting in a significant way their amplitude.

Figure 2 depicts a comparison between  $b_r(\eta=0,t)$ as obtained from  the solution of the full nonlinear reduced set of MHD equations (1)-(6) (full line), and the solution of eq. (9) (dashed line), both for $\delta =0.0052$. For that value of $\delta$ the growth rate is $\gamma = 0.1$.  Thus, even though $\delta \ll 1$ the characteristic time for saturation is quite realistic at $\sim 7$ orbital periods.  The two solutions practically overlap during the first few periods of the nonlinear oscillations. The phase difference between the two solutions is noticeable only later on due to the accumulating effect of a slight difference in the period. The latter, as has been discussed above, depends on and is very sensitive to the initial conditions. The amplitude of the nonlinear oscillations, which signifies the saturation level of the MRI, is the same for both calculations for the entire calculation time. The growth immediately after each minimum value corresponds to exponential growth with the corresponding $\gamma$ value, i.e., $0.1$.
As a demonstration of the mode interaction mechanism that results in the saturation of the MRI, Fig. 3 depicts the perturbation in the number density at the equatorial plane, as obtained from the solution of eqs. (1)-(6) for $\delta =0.0052$. The periodic energy transfer from the unstable MRI $k=1$ mode to the corresponding driven MS wave is indeed apparent.

\textit{Conclusions - } A novel non dissipative mechanism for the saturation of the MRI has been proposed by which the latter drives non resonantly MS waves in a bursty oscillatory manner. A second order Duffing equation is derived for the amplitude of the MRI. The cubic term in that equation reflects the saturation mechanism in which the driven MS waves modify the plasma beta in a periodic way. Since the system is only slightly supercritical, due to that modification the average modified plasma beta oscillates above and below the instability threshold and thus instigates the observed bursty nonlinear oscillations. Analytical solutions of the proposed Duffing equation coincide with the solutions obtained from the numerical simulations of the reduced nonlinear MHD thin disk equations. Furthermore, the same dynamical behavior repeats itself near all threshold points $\beta_{cr}^k$, while numerical simulations indicate that the bursty oscillations persist also far away from $\beta_{cr}^1$  \cite{7}. Therefore, the new saturation mechanism presented in the current work imposes severe limitations on the efficiency of the MRI to directly generate significant levels of turbulence in thin accretion disks.

%
%
\textit{Acknowledgment -} This work was supported by grant number 180/10 of the Israel Science Foundation.

\end{document}